\documentclass[twocolumn,aps,prl,showpacs]{revtex4}
\begin{document}
\title{Bell's Theorem without Inequalities and without Alignments}
\author{Ad\'{a}n Cabello}
\email{adan@us.es}
\affiliation{Departamento de F\'{\i}sica
Aplicada II, Universidad de Sevilla, 41012 Sevilla, Spain}
\date{\today}
%First version: February 17, 2003.
%This version: November 25, 2003.
%After PRL proofs.
%Minor corrections: February 12, 2004.

%%%%%%%%%%%%%%%%%%%%%%%%%%%% Abstract %%%%%%%%%%%%%%%%%%%%%%%%%%%%

\begin{abstract}
A proof of Bell's theorem without inequalities is presented which
exhibits three remarkable properties: (a) reduced local states are
immune to collective decoherence; (b) distant local setups do not
need to be aligned, since the required perfect correlations are
achieved for any local rotation of the local setups; (c) local
measurements require only individual measurements on the qubits.
Indeed, it is shown that this proof is essentially the only one
which fulfils (a), (b), and (c).
\end{abstract}

%%%%%%%%%%%%%%%%%%%%%%%%%%%%%%%%%%%%%%%%%%%%%%%%%%%%%%%%%%%%%%%%%%%

%Quantum communication
\maketitle

%%%%%%%%%%%%%%%%%%%%%%%%%%%%%%%%%%%%%%%%%%%%%%%%%%%%%%%%%%%%%%%%%%%

The proofs of Bell's theorem without
inequalities~\cite{GHZ89,Mermin90c,Hardy92,Hardy93,Cabello01a,Cabello01b}
are based on the existence, predicted by quantum mechanics, of
certain {\em perfect} correlations between results of spacelike
separated measurements. However, perfect (or almost perfect)
correlations between results of distant measurements are difficult
to achieve in real experiments~\cite{BBDH97}. Besides
``practical'' reasons such as imperfect preparations or imperfect
detector efficiencies~\cite{efficiency}, there are two main
difficulties for obtaining perfect correlations between distant
measurements. The first is decoherence, i.e., the fact that
reduced quantum states suffer unwanted couplings with the
environment during their flight to the distant regions. The second
is the need for a {\em perfect} alignment between the source
emitting entangled states and the setups of the distant
measurements~\cite{Neeman}. For the
proofs~\cite{GHZ89,Mermin90c,Hardy92,Hardy93,Cabello01a,Cabello01b},
any imperfection in the required alignments leads to the
disappearance of the required perfect correlations.

In this Letter it is shown that both difficulties can be overcome.
For this purpose, a proof of Bell's theorem without inequalities
for two observers is introduced. This proof exhibits three
remarkable properties: (a) reduced local states are immune to
collective decoherence, (b) distant local setups do not need to be
aligned, since the required perfect correlations are achieved for
{\em any} local rotation of the local setups, and (c) local
measurements require only individual measurements on the qubits.
Property (c) is very useful for practical purposes because, as
shown below, in order to fulfil (a) and (b), each of the two local
subsystems should consist of at least four qubits. Indeed, it will
be shown that the proposed proof is essentially the only one which
fulfils (a), (b), and (c).

We shall assume that, during their flight, the reduced quantum
states suffer a particularly relevant form of decoherence known as
collective decoherence~\cite{DG97,ZR97a,LCW98,KBLW01}. Collective
decoherence occurs whenever the spatial/temporal separation
between the qubits is small relative to the correlation
length/time of the environment. In this scenario, the environment
couples with the qubits without distinguishing between them and,
as a consequence of the interaction, all qubits undergo the same
unknown but unitary evolution. Therefore, a state~$|\psi\rangle$
of~$N$ qubits is immune to collective decoherence if and only
if~$|\psi\rangle$ is invariant under the tensor product of~$N$
equal unitary operators, i.e., $U^{\bigotimes N} |\psi\rangle=
|\psi\rangle$~\cite{DG97,ZR97a,LCW98,KBLW01}. States of this type
exist for~$N$ even and the smallest nontrivial subspace spanned by
such states occurs for~$N=4$
qubits~\cite{DG97,ZR97a,LCW98,KBLW01}.

Strategies to establish a common direction or Cartesian frame
between distant observers to any desired accuracy have attracted
much attention in recent
times~\cite{PS01a,PS01b,BBBMT01,BBM01a,BBM01b}. These papers have
drawn attention to two points relevant to our discussion. First,
a shared common reference frame so that distant observers may
prepare and measure spin components relative to it should not be
considered a free preexisting element in any communication
scenario but should instead be considered an expensive resource.
Second, if such a resource is not given, establishing a perfect
alignment between local reference frames requires an infinite
amount of communication. This has motivated the interest in
methods for the distribution of quantum information between
parties who do not share any reference
frame~\cite{BRS03,BEGKCW03}.

The proposed proof of Bell's theorem without inequalities is as
follows. Consider a source emitting systems of eight qubits
prepared in the state
\begin{equation}
|\eta\rangle =
(|\phi_0\phi_0\rangle+\sqrt{3}|\phi_0\phi_1\rangle+\sqrt{3}|\phi_1\phi_0\rangle)/\sqrt{7},
\label{H1}
\end{equation}
where~$|\phi_0\rangle$ and~$|\phi_1\rangle$ are the two singlet
states obtained adding up four spin-$\frac{1}{2}$ momenta,
\begin{eqnarray}
|\phi_0\rangle & = & {1 \over 2}
(|0101\rangle-|0110\rangle-|1001\rangle+|1010\rangle), \label{P0}
\\
|\phi_1\rangle & = & {1 \over 2 \sqrt{3}}
(2|0011\rangle-|0101\rangle-|0110\rangle-|1001\rangle \nonumber \\ & &
-|1010\rangle+2|1100\rangle). \label{P1}
\end{eqnarray}
These states were introduced by Kempe {\em et al.\ }in the context
of decoherence-free fault-tolerant universal quantum
computation~\cite{KBLW01}. Let us suppose that the first four
qubits prepared in~$|\eta\rangle$ fly to Alice and the second four
qubits fly to a distant observer, Bob. On her/his four qubits,
each observer randomly chooses to measure either~$F$ or~$G$,
defined as
\begin{eqnarray}
F & = & -|\phi_0\rangle\langle\phi_0|+|\phi_1\rangle\langle\phi_1|, \\
G & = & -|\psi_0\rangle\langle\psi_0|+|\psi_1\rangle\langle\psi_1|,
\end{eqnarray}
where~$|\psi_0\rangle$ and~$|\psi_1\rangle$ are obtained,
respectively, from~$|\phi_0\rangle$ and~$|\phi_1\rangle$, by
permuting qubits~2 and~3, i.e.,
\begin{eqnarray}
|\psi_0\rangle & = & {1 \over 2} (|0011\rangle-|0110\rangle-|1001\rangle+|1100\rangle) \nonumber \\
& = & {1 \over 2} \left(|\phi_0 \rangle + \sqrt{3} |\phi_1\rangle\right), \label{S0} \\
|\psi_1\rangle & = & {1 \over 2 \sqrt{3}}
(-|0011\rangle+2|0101\rangle-|0110\rangle-|1001\rangle \nonumber \\
& & +2|1010\rangle-|1100\rangle) \nonumber \\
& = & {1 \over 2} \left( \sqrt{3} |\phi_0 \rangle - |\phi_1 \rangle\right). \label{S1}
\end{eqnarray}
The observable~$F$ ($G$) has three possible outcomes:~$-1$,
corresponding to~$|\phi_0\rangle$ ($|\psi_0\rangle$), $1$
corresponding to~$|\phi_1\rangle$ ($|\psi_1\rangle$), and~$0$,
which never occurs because the local subsystems have total spin
zero. Measuring~$F$ is thus equivalent to distinguishing with
certainty between~$|\phi_0\rangle$ and~$|\phi_1\rangle$ with a
single test on the four qubits, and measuring~$G$ is equivalent to
distinguishing with certainty between~$|\psi_0\rangle$
and~$|\psi_1\rangle$. Alice's measurements on qubits~1 to~4 are
assumed to be spacelike separated from Bob's measurements on
qubits~5 to~8.

The state~$|\eta\rangle$ can also be expressed as
\begin{eqnarray}
|\eta\rangle & = & (4
|\phi_0\psi_0\rangle+\sqrt{3}|\phi_1\psi_0\rangle+3|\phi_1\psi_1\rangle)/2\sqrt{7}
\label{H2} \\ & = & (4
|\psi_0\phi_0\rangle+\sqrt{3}|\psi_0\phi_1\rangle+3|\psi_1\phi_1\rangle)/2\sqrt{7}
\label{H3} \\ & = & (7
|\psi_0\psi_0\rangle+3\sqrt{3}|\psi_0\psi_1\rangle+3\sqrt{3}|\psi_1\psi_0\rangle
\nonumber \\ & & -3|\psi_1\psi_1\rangle)/4\sqrt{7}. \label{H4}
\end{eqnarray}
Moreover, since~$|\phi_0\rangle$, $|\phi_1\rangle$,
$|\psi_0\rangle$, and~$|\psi_1\rangle$ are invariant under the
tensor product of four equal unitary operators, then they are
invariant under local rotations. Therefore,
expressions~(\ref{H1}) and (\ref{H2})--(\ref{H4})
remain unchanged after local rotations. Consequently, if~$R_A$
and~${\cal R}_A$ ($R_B$ and~${\cal R}_B$) are rotations of Alice's
(Bob's) setups for measuring, respectively, $F$ and~$G$ relative
to the reference frame of the source then, in the
state~$|\eta\rangle$, for {\em any} rotations~$R_A$, ${\cal R}_A$,
$R_B$, and~${\cal R}_B$,
\begin{eqnarray}
P(R_A F =1,R_B F=1) & = & 0, \label{G4} \\
P(R_A F =1\,|\,{\cal R}_B G=1) & = & 1, \label{G3} \\
P(R_B F =1\,|\,{\cal R}_A G=1) & = & 1, \label{G2} \\
P({\cal R}_A G=1,{\cal R}_B G=1) & = & {9 \over 112}, \label{G1}
\end{eqnarray}
where~$P(R_A F =1,R_B F=1)$ is the joint probability that both
Alice and Bob obtain the outcome~$1$ when both perform
experiment~$F$ (or any experiment consisting on independently
rotating their setups for measuring~$F$), and~$P(R_A F
=1\,|\,{\cal R}_B G=1)$ is the probability that Alice obtains the
outcome~$1$ when she performs experiment~$F$ (or any experiment
consisting on rotating her setup for measuring~$F$), conditioned
to Bob obtaining the outcome~$1$ when he performs experiment~$G$
(or any experiment consisting on rotating his setup for
measuring~$G$).

From property (\ref{G1}), if both Alice and Bob choose the setup
for measuring~$G$, then in~$8\%$ of the events the outcome is~$1$
in both cases. This is true even if Alice applies any
rotation~${\cal R}_A$ to her setup and Bob applies any
rotation~${\cal R}_B$ to his setup.

From property (\ref{G2}), if Alice measures~$G$ and obtains the
outcome~$1$, then she can predict with certainty that, if Bob
measures~$F$, he will obtain~$1$. According to Einstein, Podolsky,
and Rosen (EPR), this fact must be interpreted as sufficient
evidence that there is a local ``element of reality'' in Bob's
qubits determining this outcome~\cite{EPR35}. Moreover, EPR
reasoning seems to be even more inescapable in our example, since
Alice's prediction with certainty is valid even if Alice applies
any rotation~${\cal R}_A$ to her setup for measuring~$G$ and Bob
applies any rotation~$R_B$ to his setup for measuring~$F$.

Analogously, from property (\ref{G3}), if Bob measures~$G$
(or~${\cal R}_A G$) and obtains~$1$, then he can predict with
certainty that, if Alice measures~$F$ (or~$R_A F$), she will
obtain~$1$. Again, according to EPR, there must be a local element
of reality in Alice's qubits determining this outcome.

Therefore, assuming EPR's point of view, for at least~$8\%$ of the
systems prepared in the state~$|\eta\rangle$, there must be two
joint local elements of reality: one for Alice's qubits,
corresponding to~$R_A F=1$, and one for Bob's qubits,
corresponding to~$R_B F=1$. However, this inference is in
contradiction with property (\ref{G4}), which states that the
joint probability of obtaining the outcomes~$R_A F = 1$ and~$R_B
F= 1$ is zero. This provides a simple and powerful proof that the
concept of element of reality, as defined by EPR, is incompatible
with quantum mechanics, even if the predictions with certainty are
valid not only for a particular alignment of the distant setup but
for {\em any} possible alignment.

The logical argument in the previous proof is similar to the one
in Hardy's~\cite{Hardy93}. However, this proof exhibits some
remarkable features:

{\em (a) Partial states are immune to collective
decoherence.}---While the reduced states required in previous
proofs of Bell's theorem without
inequalities~\cite{GHZ89,Mermin90c,Hardy92,Hardy93,Cabello01a,Cabello01b}
are destroyed under collective decoherence, the reduced states
used in the proof above are immune to collective decoherence. This
can be seen by expressing the reduced density matrix describing
both local states as
\begin{equation}
\rho = [ ( 7+\sqrt{13} ) |\chi^+\rangle \langle \chi^+|+ (
7-\sqrt{13} ) |\chi^-\rangle \langle \chi^-| ]/14,
\end{equation}
where
\begin{equation}
| \chi^\pm \rangle = [ ( 1 \pm \sqrt{13} ) |\phi_0\rangle + 2
\sqrt{3} |\phi_1\rangle ] / \sqrt{26 \pm 2 \sqrt{13}}.
\end{equation}
Since~$|\phi_0\rangle$ and~$|\phi_1\rangle$ are invariant under
any tensor product of four equal unitary operators, then any
incoherent superposition of them, such as~$\rho$, is also invariant
and therefore is immune to collective decoherence.

{\em (b) Distant local setups do not need to be aligned, since the
required perfect correlations are achieved for any local rotation
of the setups.}---This property is derived from the fact that
measuring the local observable~$F$ ($G$) is equivalent to
distinguishing with certainty between the orthogonal
states~$|\phi_0\rangle$ and~$|\phi_1\rangle$ ($|\psi_0\rangle$
and~$|\psi_1\rangle$), and that both states (and thus any other
two states obtained by permuting qubits) are invariant under any
tensor product of four equal unitary operators, and thus under any
local rotation of the setup for measuring~$F$ ($G$).

{\em (c) Local observables can be measured by means of tests on
individual qubits.}---A practical advantage and a very remarkable
property of this proof is that measuring~$F$ or~$G$ does not
require collective measurements on two or more qubits but instead
a single test on each of the four qubits. Measuring~$F$ is
equivalent to distinguishing between~$|\phi_0\rangle$
and~$|\phi_1\rangle$ with a single test. Remarkably, the only two
orthogonal states invariant under any tensor product of four equal
unitary operators that can be reliably distinguished by fixed (as
opposed to conditioned, as those in~\cite{WSHV00}) measurements on
the four individual qubits are~$|\phi_0\rangle$
and~$|\phi_1\rangle$ and those obtained from them by permuting
qubits (such as~$|\psi_0\rangle$ and~$|\psi_1\rangle$). To prove
this, let us consider two orthogonal states invariant under any
tensor product of four equal unitary operators,
\begin{eqnarray}
|\psi \rangle & = & \cos \omega |\phi_0 \rangle + \sin \omega
|\phi_1 \rangle, \\
|\psi^{\perp} \rangle & = & \sin \omega |\phi_0 \rangle - \cos
\omega |\phi_1 \rangle.
\end{eqnarray}
States~$|\psi \rangle$ and~$|\psi^{\perp} \rangle$ are reliably
distinguishable by fixed measurements on the four individual
qubits if and only if there is an orthogonal local
basis~$\{|0_a\rangle \otimes |0_b\rangle \otimes |0_c\rangle
\otimes |0_d\rangle, \ldots,|1_a\rangle \otimes |1_b\rangle
\otimes |1_c\rangle \otimes |1_d\rangle\}$ in which, for all~$j$
such that the~$j$ component of~$|\psi \rangle$ ($|\psi^{\perp}
\rangle$) is not zero, then the~$j$ component of~$|\psi^{\perp}
\rangle$ ($|\psi \rangle$) is necessarily zero. Since~$|\psi
\rangle$ and~$|\psi^{\perp} \rangle$ are invariant under any
tensor product of four equal unitary operators, we can restrict
our attention, without losing generality, to the case of spin
measurements in the~$x$-$z$ plane. Then, orthogonal local basis
are composed by states of the form
\begin{eqnarray}
|0_a \rangle & = & \cos {\theta_a} |0\rangle + \sin {\theta_a}
|1\rangle, \\
|1_a \rangle & = & \sin {\theta_a} |0\rangle - \cos {\theta_a}
|1\rangle.
\end{eqnarray}
If~$|\psi \rangle$ and~$|\psi^{\perp} \rangle$ are distinguishable
in a basis of this type, then, when expressed in such a basis, the
first component of one of the two states must be zero. After some
algebra, it can be seen that a necessary condition for the first
component of~$|\psi \rangle$ to be zero is
\begin{eqnarray}
\cot \omega & = & {1 \over \sqrt{3}} \csc\left({\theta_a-\theta_b}\right)
\csc\left({\theta_c-\theta_d}\right) \nonumber \\
& & \times [\cos\left({\theta_a+\theta_b-\theta_c-\theta_d}\right) \nonumber \\
& & -\cos\left({\theta_a-\theta_b}\right)
\cos\left({\theta_c-\theta_d}\right)]. \label{cot}
\end{eqnarray}
This condition makes zero both the first and last component
of~$|\psi \rangle$. Components two to eight can be expressed in
terms of~$\omega$, $\theta_a$, $\theta_b$, $\theta_c$,
and~$\theta_d$. Components nine to $15$ are identical to
components two to eight but in reverse order and with opposite
signs. If~$|\psi \rangle$ and~$|\psi^{\perp} \rangle$ are
distinguishable in a basis of this type, then more components
should also be zero. The important point is that the cost of
obtaining more zeroes is to restrict the possible values
of~$\omega$. It can be checked that any way to make more than four
zeroes (one of the two orthogonal states must have more than four
zeroes) requires
\begin{equation}
\omega = n \pi / 6
\end{equation}
(with $n$ integer). We therefore conclude that the only two
four-qubit states invariant under any tensor product of four equal
unitary operators that are reliably distinguishable by fixed
measurements on individual qubits are those with~$\omega = n \pi /
6$ (with $n$ integer). Note, however, that these states are
precisely those obtained from~$|\phi_0 \rangle$ and~$|\phi_1
\rangle$ by permuting qubits.

We shall now show that, to distinguish with certainty
between~$|\phi_0 \rangle$ and~$|\phi_1 \rangle$, it is enough to
measure the spin component of the first two qubits along the same
direction and the spin component of the other two qubits along a
perpendicular direction. This can be seen by resorting to the
invariance under any tensor product of four equal unitary
operators and expressing these states in the basis of eigenstates
of~$\sigma_{z1} \otimes \sigma_{z2} \otimes \sigma_{x3} \otimes
\sigma_{x4}$,
\begin{eqnarray}
|\phi_0 \rangle & = & {1 \over 2}
(-|01\bar{0}\bar{1}\rangle+|01\bar{1}\bar{0}\rangle+|10\bar{0}\bar{1}\rangle-|10\bar{1}\bar{0}\rangle),
\label{lp0}
\\
|\phi_1 \rangle & = & {1 \over 2 \sqrt{3}}
(|00\bar{0}\bar{0}\rangle-|00\bar{0}\bar{1}\rangle-|00\bar{1}\bar{0}\rangle+|00\bar{1}\bar{1}\rangle \nonumber \\
& & -|01\bar{0}\bar{0}\rangle+|01\bar{1}\bar{1}\rangle
-|10\bar{0}\bar{0}\rangle+|10\bar{1}\bar{1}\rangle \nonumber \\ &
&
+|11\bar{0}\bar{0}\rangle+|11\bar{0}\bar{1}\rangle+|11\bar{1}\bar{0}\rangle+|11\bar{1}\bar{1}\rangle),
\label{lp1}
\end{eqnarray}
where~$\sigma_z |0\rangle=|0\rangle$, $\sigma_z
|1\rangle=-|1\rangle$, $\sigma_x |\bar{0}\rangle=|\bar{0}\rangle$,
$\sigma_x |\bar{1}\rangle=-|\bar{1}\rangle$ [$|\bar{0}\rangle
=(|0\rangle+|1\rangle)/\sqrt{2}$ and~$|\bar{1}\rangle
=(|0\rangle-|1\rangle)/\sqrt{2}$]. According to~(\ref{lp0})
and~(\ref{lp1}), if the measurements on the individual qubits
are~$\sigma_{z1}$, $\sigma_{z2}$, $\sigma_{x3}$, $\sigma_{x4}$ (or
any rotation thereof), then, among the $16$ possible outcomes,
four occur (with equal probability) only in the state~$|\phi_0
\rangle$, and the other twelve occur (with equal probability) only
in the state~$|\phi_1 \rangle$ (this has been experimentally
demonstrated in~\cite{BEGKCW03}). Therefore, to measure~$F$ ($G$),
it is enough to measure the spin component of qubits~$1$ and~$2$
($1$ and~$3$) along the same direction and the spin component of
the other two qubits along a perpendicular direction.

{\em (d) Contradiction is nearly optimal.}---The fact that, except
for permutations of the qubits, the only two orthogonal four-qubit
states invariant under any tensor product of four equal unitary
operators that can be reliably distinguished by fixed measurements
on individual qubits are~$|\phi_0\rangle$ and~$|\phi_1\rangle$
means that the only local observables whose eigenvectors are
invariant under any tensor product of four equal unitary operators
and can be reliably distinguished by fixed measurements on the
four individual qubits are, precisely, $F$ and $G$. This
enormously restricts the possible proofs of Bell's theorem without
inequalities which satisfy (a), (b), and (c). By checking every
possible combination of states and observables, it can be seen
that the proof presented here exhibits the maximum probability,
${9 \over 112} \approx 0.08$, for a Hardy-like contradiction
satisfying (a), (b), and~(c). Without requirement (c), it can be
proven (as in~\cite{Hardy93}) that the maximum probability for a
Hardy-like contradiction is~$[(\sqrt{5}-1)/2]^5 \approx 0.09$.
Therefore, the maximum probability for a Hardy-like contradiction
satisfying (a), (b), and~(c) is close to the optimal probability
without these requirements.

In summary, some recent methods of decoherence-free fault-tolerant
universal quantum computation have been used to illustrate that
some difficulties that were previously assumed to be inherent to
any proof of Bell's theorem without inequalities can indeed be
overcome.

On the experimental side, while the four-qubit states~$|\phi_0
\rangle$ and~$|\phi_1 \rangle$ have already been prepared and
their immunity to collective decoherence and invariance under
local rotations tested in a laboratory~\cite{BEGKCW03}, preparing
entangled superpositions thereof, such as~$|\eta\rangle$, is a
significant goal for future research.

%%%%%%%%%%%%%%%%%%%%%%%%%%%%%%%%%%%%%%%%%%%%%%%%%%%%%%%%%%%%%%%%%%%

I thank M. Bourennane and H. Weinfurter for useful discussions, J.
L. Cereceda and N. D. Mermin for their comments on the manuscript,
and the Spanish Ministerio de Ciencia y Tecnolog\'{\i}a Project
No.~BFM2002-02815, the Junta de Andaluc\'{\i}a Project
No.~FQM-239, and the European Science Foundation (Quantum
Information Theory and Quantum Computation Short Scientific Visit
Grants Program) for support.

%%%%%%%%%%%%%%%%%%%%%%%%% References %%%%%%%%%%%%%%%%%%%%%%%%%


\begin{thebibliography}{99}
\bibitem{Neeman}
For instance, Yuval Ne'eman defended that the answer to the puzzle
posed by Bell's theorem was to be found in the implicit assumption
that the detectors were aligned. Ne'eman apparently believed that
the two detectors were connected through the space-time affine
connection of general relativity [N.D. Mermin (private
communication)].
\end{thebibliography}
\end{document}